# Processing of Electronic Health Records using Deep Learning: A review

**Venet Osmani**
e-health Research Group
Fondazione Bruno Kessler
Trento, Italy
vosmani@fbk.eu

**Li Li, Matteo Danieletto, Benjamin Glicksberg, Joel Dudley**
Icahn School of Medicine at Mount Sinai
Institute of Next Generation of Healthcare
New York, NY 10065, USA
{li.li, matteo.danieletto, benjamin.glicksberg, joel.dudley }@mssm.edu

**Oscar Mayora**
e-health Research Group
Fondazione Bruno Kessler
Trento, Italy
omayora@fbk.eu

## ABSTRACT

*Background*. Availability of large amount of clinical data is opening up new research avenues in a number of fields. An exciting field in this respect is healthcare, where secondary use of healthcare data is beginning to revolutionize healthcare. Except for availability of Big Data, both medical data from healthcare institutions (such as EMR data) and data generated from health and wellbeing devices (such as personal trackers), a significant contribution to this trend is also being made by recent advances on machine learning, specifically deep learning algorithms.

*Objectives*. The objective of this work was to provide an overview of how automatic processing of Electronic Medical Records (EMR) data using Deep Learning techniques is contributing to understating of evolution of chronic diseases and prediction of risk of developing these diseases and associated complications.

*Methods*. A review of the scientific literature was conducted using scientific databases Google Scholar, PubMed, IEEE, and ACM. Searches were focused on publications containing terms related to both Electronic Medical Records and Deep Learning and their synonyms.

*Results*. The review has shown that a number of studies have reported results that provide unprecedented insights into chronic diseases through the use of deep learning methods to analyze EMR data. However, a major roadblock that may limit how effectively these paradigms can be utilized and adopted into clinical practice is in the interpretability of these models by medical professionals for whom many of them are designed.

*Conclusions*. Despite the identified challenges automatic processing of EMR data with state-of-the-art machine learning approaches, such as deep learning, will push predictive power well beyond the current success rates. Hopefully, we will continue to see findings from these works to continue to transform clinical practices, leading to more cost effective and efficient hospital systems along with better patient outcomes and satisfaction.

## 1 INTRODUCTION

Health care systems across the world are adopting Electronic Medical Records (EMR), providing a collection of longitudinal data pertaining to patients' health. Use of electronic records will significantly increase the availability of clinical data as well as impact on the potential of discovering of new disease patterns as well as providing personalized patient care by automatically processing of this vast quantity of data. Considering that richness of information contained in EMRs [1] and their potential to transform delivery of care [2], understanding the information contained in EMRs is becoming an important challenge. In this respect the rise of Deep Learning is accelerating automatic processing and understanding of EMR data. Especially the use of novel Deep Learning methods and architectures that can handle multi-dimensional, heterogenous and incomplete data are seen as particularly promising. These developments are foreseen to provide an increasing uptake in precision medicine through development of personalized health services enabled by analysis of individual and aggregated multimodal data residing into patients' Electronic Medical Records (EMR). In this paper we provide a review of the combination of these technologies and their impact on the delivery of care.

## 2 METHODS

We conducted and reported the review according to the Preferred Reporting Items for Systematic Reviews and

Meta-Analysis (PRISMA) statement. Methods of the review process and eligibility criteria were established in and remained unchanged during the review. We have conducted a literature search using electronic databases Google Scholar, PubMed, IEEE, and ACM. The literature search was supplemented by manual retrieval of references contained in the articles. The literature search was conducted without time restrictions, using the following keywords: (EMR OR EHR OR PHR OR "Electronic Medical Records" OR "Electronic Health Records" OR "Personal Health Records") and ("Deep Learning"). Majority of articles covered a period from 2010 (beginning of the popularization of Deep Learning) up to 2017. Our search strategy retrieved 1790 articles. All identified titles and abstracts were screened for eligibility by 2 researchers (VO and LL). Articles were excluded based on two criteria: 1) They did not primarily focus on processing of health records using deep learning or 2) the work did not make use of Deep Learning methods. After this screening, there were 169 full text articles that were further assessed for relevance. After this screening, there were 36 articles that were considered for this review.

## 3   RESULTS

Out of 1790 articles identified in the initial search, there were 36 articles that fulfilled eligibility criteria and were selected for review. We have divided the studies into two categories namely i) disease modelling using EMR data and ii) an overview of Deep Learning architectures and their use with EMR data.

### 3.1   Building disease models

Broadly, EMR are the collection of healthcare related data captured during a patient visit to a hospital or clinic, including disease diagnoses, medication prescriptions, procedures, surgeries, and lab test results among others. One of the most powerful applications of EMR data is for the refinement of personalized medicine strategies [3], as it provides an unparalleled amount of population-sized, patient-level data that can be mined to directly inform clinical practice. In fact, Precision Medicine Initiative (https://www.whitehouse.gov/precision-medicine) was launched with a $215 million investment to enable clinical practice to move from a "one-size-fits-all approach" to more individualized care. When EMR data is combined with other modalities of data such as molecular features, biosensors, social determinants of health, and environmental exposures among others, a rich opportunity is created to study an individual patient's disease in multiple dimensions and at multiple scales. Initiatives such as the personalized cancer therapy program [4] have utilized EMR resources to tailor treatment regimens to participating patients. Moreover, additional data parameters from EMR are used to augment current traditional Modified Early Warning Score (MEWS) algorithms, which "track-and-"trigger" warnings of patient condition deterioration based on six cardinal vital signs [5], especially when combined with continuous monitoring of patients [6-10] Furthermore, discoveries from data-driven, EMR-based research can lead to actionable findings, such as identifying medication adverse reactions [11] [12] and predicting future disease risk [13]. In addition to enabling the provision of clinical care, EMR are also a powerful tool to assist fundamental research [14]. There are countless examples of research studies that discovered patterns of disease susceptibility [15, 16], comorbidity [17] [18], and trajectories [19, 20] using EMR. Linking EMR to other -omics data types within a network biology framework [21] has helped to elucidate contributions of various risk factors to disease etiology, including genetics [22] [23], environment [24], demography [25, 26], and combinations thereof [27]. Genetic and genomic data are often linked with clinical data via hospital-affiliated biobanks, which recruit individuals from the visiting patient population to obtain and "bank" their genetic data. This approach has led to the development of fields of research such as phenome-wide association studies (PheWAS) [28], where a traditional genome-wide association study (GWAS) analysis is performed on a large retrospective hospital cohort allowing more freedom in phenotype selection. Recently, using whole-exome sequencing and a linked EMR system for over 50,000 participants, the DiscovEHR study revealed the clinical impact and phenotypic consequences of functional variants [29]. Adding in medication RNA expression signatures further facilitates pharmacological discoveries, ranging from drug discovery [30] and repurposing [31-33] to pharmacogenomics [34, 35]. In the next subsections, we provide two examples of disease modeling, namely diabetes and mental disorders.

#### 3.1.1   Using topological data analysis to characterize type 2 diabetes mellitus subgroups

Type 2 diabetes mellitus (T2D) is highly complex, heterogeneous, and multifaceted disease that stems from an intricate interplay between genetic background and environmental factors [24]. While a diagnosis of T2D is diagnosed using a standardized assessment of blood glucose measures (e.g. hemoglobin A1c and/or fasting plasma glucose), it is clear that the etiological landscape of the disorder is mechanistically diverse [36]. Clinical presentation of the disease often varies and treatment is often hard for patients to manage [37]. For these reasons and others, more personalized treatment styles need to be developed to facilitate better disease management.

With this in mind, researchers developed an elaborate data-driven, unsupervised analysis coupling genetic and clinical data from EMR in order to better characterize the T2D disease landscape [38]. They utilized the Charles Bronfman Institute of Personalized Medicine's BioMe biobank as a research cohort, which is a genotyped subset of patients from Mount Sinai Hospital with linked medical records (n=11,210). For all patients, they first only considered hospital encounters occurring within 30 days of study enrollment, compiling all available laboratory test results, disease diagnoses (in the form of ICD-9 codes, but organized via Clinical Classifications Software [39], and medication prescriptions.

To ensure high quality analysis, they used the robust T2D eMERGE electronic phenotyping algorithm [40, 41] (https://phekb.org/phenotype/); which details specific inclusion/exclusion criteria for multiple data types) to designate cases and controls for the study. This algorithm is open-source and can be obtained within the repository. Briefly, the groups are defined through multiple inclusion and exclusion criteria across multiple modes of information. For instance, individuals without T2D diagnosis, but who have an abnormal plasma glucose measurement are excluded from the control group. As much of the data is sparse, they only incorporated variables in which over half the patients had values for (n=73) in the actual analysis.

Once the patient/control cohorts were designated, they performed topological data analysis (TDA)-based unsupervised clustering on the included clinical variables to identify any potential substructure of T2D using Ayasdi (http://ayasdi.com, Ayasdi Inc.). Additionally, for all data points, they calculated a cosine similarity metric to assess relatedness of clinical features. Next, they applied two filter functions, (1) L-infinity centrality and (2) principal metric singular value decomposition to produce a patient-patient network. In this network, distance between nodes (i.e. distance between patients) reflects similarity of clinical feature composition.

From the T2D topological network, they were able to identify three robust clusters of patients, which implicate potential subtypes of the disease. For the patients within each cluster, they compared genetic composition (i.e. SNP prevalence and pathway enrichment), disease comorbidities (via CCS classification), and clinical traits such as platelet counts. They discovered interesting genotype and phenotype disparities between clusters that begin to disentangle the extreme heterogeneity of the disease. For instance, patients in subtype 2 (~24% of all T2D cases) had the lowest weight, a unique genetic background (1,227 uniquely associated single nucleotide variants in 322 genes), and were enriched for cancer malignancy (e.g. cancer of bronchus and lung). These individuals may constitute the established cancer/T2D epidemiological connection [38] . This approach is an example of disease stratification. With these findings, they demonstrated that highly complex chronic diseases could be deconstructed using EMR and affiliated biobank data, which can both augment disease reclassification as well as precision medicine efficiency.

### 3.1.2 Advancing mental illness prediction using deep learning

Another compelling use of EMR-based research is the prediction of future disease outcomes in a data-driven approach using rich longitudinal patient data. Many chronic conditions are particularly difficult for the patient to manage and end up requiring large costs by healthcare systems due to continual re-hospitalizations. Mental illnesses, such as schizophrenia or attention deficit hyperactivity disorder (ADHD), are a heterogeneous group of conditions that are difficult to treat due to the interplay between psychological, psychiatric, and neurological aspects and inherent heterogeneity of symptoms, unique presentations, and somewhat imprecise and subjective diagnostic criteria. Additionally, socioeconomic and environmental factors [42], such as access to healthcare or social pressure, may prevent individuals from getting treatment or even knowing that they have a condition [43]. Therefore, continuous sensing of patients' state in an unobtrusive manner [44, 45] such as through fitness trackers or smartphone sensors [46-48] provides valuable information in identifying early warning signs and serve as an aid in identifying most appropriate treatment plans.

Recently, researchers developed a deep learning framework applied to a large-scale EMR database called Deep Patient [49] to construct predictive models for diseases across multiple domains, including mental illness. Using the EMR data, they collected all available clinical data (e.g. lab tests, etc.) for patients (n=704,587) with at least one disease diagnosis as well as five other recorded encounters up to 2014. They then further filtered clinical descriptors, removing those that appeared in >80% and <5% of patients. A random subset of these patients (n=200,000) was used as training data and a separate subset (n=81,284) was used for testing (n=76,284 for test, and n=5,000 for further feature validation). The model was trained using all data up to and including 2013, and performance was then assessed by actual outcomes manifesting in 2014. They used Stacked denoising autoencoders (SDAs), which map features to hidden representations of the data at each of the three layers used, to learn general deep patient representations of all available features across all phenotypes. For all 78 assessed diseases, Deep Patient outperformed all other modeling strategies with a mean area under the ROC curve (AUC-ROC) of 0.773 (next highest was independent component analysis with 0.695), emphasizing the utility of deep learning in chronic disease risk modeling. It is clear, however, that Deep Patient (along with other patient representation models) has more predictive power for certain diseases than others, which may result from inherent heterogeneity and other factors not contained within the EMR data. With Deep Patient for instance, prediction of diabetes mellitus with complications performed the best and disorders of lipid metabolism the worst with AUC-ROCs of 0.907 and 0.561, respectively. Mental illness-related diseases tended to have reasonable predictive power with all having AUC-ROCs over 0.6 and most (6/8) over 0.75. Attention-deficit and disruptive behavior disorders performed best and anxiety disorders worst with AUC-ROCs of 0.863 and 0.605 respectively.

Deep Patient was able to quantify disease risk at the individual level, by assessing the proportion of the top one, three, and five predicted diseases per patient that appeared in different intervals within the test window (e.g. 30, 60, 90, and 120 days). Deep patient outperformed all other patient representation models in all aspects. As expected, predictions for the single top disease were more accurate than when considering the top three or five. Prediction accuracy was also positively correlated with time window length, with the most accurate predictions occurring for the 180-day window.

### 3.2 Deep Learning architectures

Automatic analysis of EMR data has been carried out using traditional machine learning methods. However, a number of issues, including characteristics of EMR data (such as heterogeneity, high-dimensionality, sparseness) have created scalability challenges, that is ability to automatically analyze large amount with low manual intervention (such as feature extraction for example). As such deep learning is particularly well positioned to analyze healthcare data, especially data contained in EMR. Deep learning is a method of machine learning that uses layers of models to predict outcomes. More specifically, it is a hierarchical neural network architecture that evaluates an input through a cascade of multiple layers. Each layer is composed of models, which compute upon data provided to it. The results of this computation, which are transformed versions of the original input, are then passed to another, higher-level model, continuing in this way until the top-level predictive model is reached, resulting in a composite model with abstracted combinations of features. Deep learning is increasingly being applied in the field of bioinformatics, such as drug discovery and, given a large enough training set, has outperformed other machine learning algorithms for predictive modeling of risk within EMR systems [40].

Traditional machine learning approaches have limited ability to process data in their raw form, requiring considerable expertise in designing feature extraction and selection. In contrast, deep learning methods learn the optimal features directly from the data, allowing automatic discovery of the representations required for classification. This is an especially important aspect when considering that EMR data is typically noisy, sparse, and incomplete as reported in [39]. In this respect, there have been several works that have used deep learning to analyze EMR data in order to better understand disease trajectories and predict risk. In terms of risk prediction Tran et. al [41] work was based on analysis of EMR data through restricted Boltzmann machines in order to provide suicide risk stratification. Work on DeepPatient used stacked autoencoders and logistic regression to predict a wide variety of disease diagnosis [39]. A number of other studies have used deep learning methods not only to predict the onset of diseases but also the likely timeframe of occurrence. In this respect Pham et. al's DeepCare framework [42] creates two vectors on each patient admission, one for diagnosis and one for intervention whereby these vectors are concatenated and provided to an LSTM network able to predict the timeframe of the next intervention for diabetes and mental health. LSTMs were used also by Lipton et. al [43] specifically a two-layer LSTM with 128 memory cells each in order to predict the most likely diagnosis out of 128 possibilities. On the other hand Cheng et. al [44] trained a CNN on a temporal matrix with time on one dimension and clinical event on the other dimension for each patient in order to predict congestive heart failure and Chronic Obstructive Pulmonary Disease on 1127 and 477 cases for each disease respectively. Having experimented with a number of CNN techniques, they found that CNN with slow fusion provided the best predictive result.

An alternative approach is presented by Choi et. al's Doctor AI [45], where the objective was to model how physicians reach decisions by predicting future diseases, along with corresponding medication. Their trained GRU performed differential diagnosis with an accuracy comparable to physicians, achieving 79% recall. It should be noted that their system showed a similar performance using a different institution's coding system, and also the performance on the publicly available MIMIC dataset [46] increased by pre-training their models on their own private data, suggesting that the network has generalized well. In all of these tasks, whenever there was a comparison with conventional machine learning approaches, it was found that deep learning approaches resulted in better performance.

## 4 CONCLUSION

There are countless studies in addition to those described above that utilize EMR systems to reveal insights into chronic diseases and transform clinical practice. It is clear, however, that there are several roadblocks that limit how effectively these paradigms can be utilized and adopted into clinical practice. One substantial limitation, for instance, is in the interpretability of these models by medical professionals for whom many of them are designed. Several utilities have been developed to facilitate full comprehension of machine learning predictions. For example EHDViz [50] (http://ehdviz.dudleylab.org/), is a clinical dashboard that visualizes data streams, in real time, from an EMR. Outcomes from predictive models can be overlaid on visual flows such as these to put them in context of actual patient data. Despite these challenges, EMR-based research will continue to evolve to produce even more outstanding insights, especially through the use of deep learning. With nomenclature standardization practices improving and resources growing, integration with developing resources of other biological and environmental modalities (i.e. pollution data) and sensor-based data collection [51-53] will allow for a multi-scale understanding of findings. Clearly genetic data will play a crucial role in this process, however considering their scarce availability, we believe automatic processing of EMR data through Deep Learning methods is a viable candidate to contribute to better understanding of chronic diseases. Hopefully, we will continue to see findings from these works in transforming clinical practices, leading to more cost effective and efficient hospital systems along with better patient outcomes and satisfaction.

## ACKNOWLEDGMENTS

This work was partially supported by the European Commission's Horizon 2020 WellCo project under grant agreement No 769765 (http://wellco-project.eu)